\documentstyle[12pt]{article}
\def\be{\begin{equation}}
\def\ee{\end{equation}}
\def\bea{\begin{eqnarray}}
\def\eea{\end{eqnarray}}
\def\a{\alpha}
\def\b{\beta}
\def\t{\tau}
\def\s{\sigma}
\def\pa{\partial}

\author{H.-J. Schmidt, U. Semmelmann}

\title{Cosmic strings and strings in gravitational waves}

\date{}
\begin{document}
\maketitle

\centerline{Universit\"at Potsdam, Institut f\"ur Mathematik, Am
Neuen Palais 10} 
 \centerline{D-14469~Potsdam, Germany,  E-mail:
 hjschmi@rz.uni-potsdam.de}

\begin{abstract}
\noindent
 We consider strings 
with the Nambu action as extremal surfaces in a given space-time, 
thus, we ignore their back 
reaction. Especially, we look for strings sharing one symmetry 
with the underlying space-time. If this 
is a non-null symmetry, the problem of determining 
the motion of the string can 
be dimensionally reduced. We get exact solutions for the 
following cases: straight and 
circle-like strings in a Friedmann background, straight 
strings in an anisotropic Kasner 
background, different types of strings in the metric 
of a gravitational wave. The solutions will be discussed.

\medskip

\noindent
Wir betrachten Strings mit der Nambu-Wirkung als Extremalfl\"achen in 
einer 
gegebenen Raum-Zeit, d.h., wir ignorieren ihre R\"uckwirkung. Wir 
interessieren uns 
dabei besonders f\"ur solche Strings, die eine Isometrie mit der
unterliegenden 
Raum-Zeit gemeinsam haben. Handelt es sich dabei um eine 
nicht-lichtartige Symmetrie, 
so l\"a\ss t sich das Problem der Bestimmung der 
Stringbewegung dimensionsreduzieren. 
Wir erhalten exakte L\"osungen f\"ur die folgenden F\"alle: gerade 
und kreisrunde 
Strings im Friedmann-Hintergrund, gerade Strings im 
an\-isotropen Kasner-Hintergrund 
sowie 
verschiedene Stringtypen in der Metrik einer Gravitationswelle. 
Die L\"osungen 
werden diskutiert.
\end{abstract}


AAA subject classification: 066

\section{Introduction}%1

To give detailed arguments for 
considering strings means carrying coals to Newcastle. So we only list 
the main points:
A string is, generally speaking, an object possessing a two-dimensional 
world surface 
(= world sheet) in contrast to a point particle possessing a 
one-dimensional world line, 
cf. the review article VILENKIN (1985). In details

1.   One considers strings with the Nambu action in a $D$-dimensional 
flat space-time. 
The theory can be consistently quantized for $D = 26$ only: otherwise 
the light cone 
quantization leads to a breakdown of Lorentz covariance, cf. 
GREEN, SCHWARZ  
and WITTEN (1987). But we consider a classical (= non-quantized) 
theory only and 
require $D = 4$ henceforth.

2.   Cosmic strings are one-dimensional topological defects in gauge 
field theories. 
They have a large mass and can be seeds for larger objects by accretion 
processes, cf. ZELDOVICH (1980).

3.   One looks for solutions of the Einstein equation with 
distribution-valued energy-momentum tensor whose support
is a two-dimensional submanifold of indefinite signature. The equation 
of state is $p_z = - \mu$ and the solution is called cosmic string. 
These are good candidates 
for seeds of galaxies in the early universe.

4.   If one looks for strings according to 2. or 3. in a given space-time, 
i.e., 
with negligible back reaction of the string onto  space-time geometry, one 
arrives at 
the Nambu action, too, see NIELSEN and OLESEN (1973) for 2. 
and GEROCH and TRASCHEN (1987) for 3. In other words, cosmic 
strings are 
extremal (i.e., maximal or minimal  in dependence of the 
boundary conditions) 
surfaces of indefinite signature in a given space-time. This approach 
is justified if 
the diameter of the string is small compared with the curvature
 radius of the underlying 
manifold and if for the string tension $\a'\ll
 1/G$  holds (we use $c = 1$). Normally, one 
thinks in orders of magnitude $G \a'=   10^{-6\pm2}$,
 see BRANDENBERGER (1987). In other 
words, the string is supposed to possess 
a mass per unit length of $10^{22\pm2}$ g/cm, if 
the phase transition is supposed to be at the GUT-scale
 (see FROLOV and SEREBRIANY 1987).

\bigskip

In FROLOV et al. (1988) a stationary  string in the Kerr-Newman metric has 
been considered. We use the method  developed there and apply it
 to other cases.

\bigskip

In the present paper we shall adopt the 4. approach and try  to give some 
geometrical 
insights into the motion of a string. To this end we give a sample 
of closed-form 
solutions for a special class of string solutions: strings which  
share a space-like 
or timelike isometry with the underlying background metric.

\bigskip

The paper is organized as follows: sct. 2 contains the main formulae, sct. 
3 the string in a Friedmann background, sct. 4 the string in the anisotropic
 Kasner background and sct. 5 the string in a gravitational wave.

\section{The main formulae}%2

The string is a two-dimensional extremal surface  of  indefinite signature. 
Let 
us take coordinates $(\t, \s) = (y^A)$, $A = 0, 1$ within the string 
and coordinates 
 $(x^i)$, $i=  0, \,  1,\,  2,\,  3$  for the space-time $V_4$ with the 
metric $g_{ij}$. The string is given 
by specifying  the
four functions $x^i(y^A)$. The signature for the metric
  $g_{ij}$  is $(+---)$. 
The tangents to the string are
\be%(1)
     e^i_A = \pa x^i/\pa y^A
\ee
and induce the metric
\be%(2)
     h_{AB} =  e^i_Ae^j_B \,  g_{ij}
\ee
at the
 string world sheet. The 
signature  of the string is required to be $(+-)$, 
thus
\be%3
     h \equiv {\rm  det} \,  h_{AB} <0 \, . 
\ee
The action to be varied is
\be%L/fflhdadr.     (4)
I = \frac{1}{2 \a ' } \int \int \sqrt{-h} \,  d\s d \t \, .
\ee
Instead of writing down the full equations we specialize to 
the case we are interested here: we require that an isometry 
for both the underlying space-time and the string exists.  
Let $k_i$  be a non-null hypersurface-orthogonal Killing vector 
field, i.e.,
\be%(5)
k_i k^i \ne 0, \quad k_{[i} k_{j;k]} =0, \quad k_{(i;j)} =0 \, .
\ee
Then there exist coordinates $x^i$ \    $(x^0 = t)$ such
\be%(6)
ds^2 = g_{00} dt^2 + g_{\a\b} dx^{\a} dx^{\b} \, .
\ee
$\a, \, \b = 1, \, 2, \,  3$,
 the $g_{ij}$
  do not depend on $t$,
 $k_i = \pa/\pa t$. 
The sign of $g_{00}$ is  determined by the 
condition $g_{00} \, k_i k^i > 0$: for timelike $k_i$
 we have $g_{00}> 0$  and for spacelike $k_i$
 we have $g_{00}< 0$. The requirement that 
$k_i$  is also an isometry of the string 
gives us the possibility to specify the functions 
$x^i(y^A)$  to be $t = \t$, $x^\a$ depends 
on $\s$ only. Then we get with (1)
$$
e^i_0 = (1, \, 0, \, 0, \, 0)\, , \quad
e^i_1 = (0, \, dx^\a/d\s) \, .
$$
 With (2, 4) we get $h_{01}=0$, $ h_{00}  = g_{00}$,
\bea%(7)
h_{11}= g_{\a\b}\  dx^{\a} /d\s \ dx^{\b} /d\s \, , \nonumber \\
I = \frac{1}{2 \a '  } \int  \int 
\sqrt{ - g_{00}\  g_{\a \b} \  dx^{\a} / d \s \ dx^{\b} / d\s }
 \ d\s d \t \, .
\eea
The integrand does not depend on $\t$, so we 
can omit the $\t$ integration. Therefore, 
extremizing (7) is the same as solving
the geodesic equation for the auxiliary metric $f_{\a\b}$
 of a $V_3$ defined by
 \be%8
f_{\a\b} = - g_{00} g_{\a\b} \, .
\ee
Remarks:  1. If $k_i$ is not hypersurface-orthogonal then eq. (6) 
contains terms with $g_{0 \a}$ and one has to add 
 $g_{0 \a} g_{0 \b}$
to the r.h.s. of eq. (8). 
2. The dimensional reduction is not fully trivial: the geodesic equation for 
(8) corresponds to compare the strings in (7) with other strings 
sharing the 
same isometry induced by $k_i$, whereas the Nambu action has 
to be compared 
with all other strings, too. But  writing down all  full
 equations or counting the degrees of freedom one can see that at our 
circumstances no difference appears.

\bigskip

For a time-like $k_i$, $f_{\a\b}$  is positive definite, 
and the condition (3) is 
automatically fulfilled. On the other hand, for a space-like $k_i$, 
$f_{\a\b}$
 is 
of signature $(+ - -)$  and eq. (3) requires the vector $dx^\a/d\s$
 to be time-like, i.e., 
the root in eq. (7) has to be real.

\section{The string in a Friedmann model}%3

Now we specify the underlying $V_4$  to be a spatially flat 
Friedmann model 
\be%9
ds^2 =dt^2 - a^2(t)(dx^2 +dy^2 +dz^2) \,  .
\ee

\subsection{The open string}%3.1.

First we use the space-like Killing vector $\pa/\pa x$ of (9). We have 
in mind
 an 
infinitely long straight string moving through the expanding universe. 
We apply 
the formalism of sct. 2 and get the following result: we write eq. (9) 
in the form 
of eq. (6)
$$
ds^2= - a^2 dx^2 + dt^2 - a^2(dy^2 + dz^2  )   \,   . 
$$
The metric (8) then reads
$$
ds^2_{(3)} = a^2 dt^2 - a^4(dy^2 + dz^2) \, .
$$
Without loss of generality the string is situated at $z = 0$  and moves into 
the $y$-direction according to 
\bea
ds^2_{(3)} = a^2 dt^2 - a^4dy^2     \nonumber \\
 \ddot t +  \frac{1}{a} \frac{da}{dt} \dot t^2 + 2
\frac{da}{dt} a \dot y^2 = 0 \quad {\rm with} \quad \cdot 
= d/d\lambda  \nonumber \\
 a^4 \dot y = M  \nonumber \\
a^2 \dot t^2 - a^4 \dot y^2 =1  \nonumber \\
y(t) = M \int \frac{dt}{a(t) \sqrt{a^4(t) + M^2}}
\eea
where $M$ is an integration constant. The natural distance of the string 
from the 
origin is
$$
s(t) = a(t) y(t) \, .
$$

\bigskip

\noindent
{\it 1.   Example.} Let $a(t) = t^n$, $ n \ge 0$
 then for $t \gg 1$, cf. STEIN-SCHABES and 
BURD (1988),
$$
     y(t)   \approx M t^{1-3n} \, , \quad s(t)
   \approx M t^{1-2n}  \quad   {\rm 
 for} \quad  n \ne 1/3
$$
and $y(t) \approx M \ln t$ for $n= 1/3$.

\bigskip

Interpretation: $   n =0$,  i.e., the absence of gravity, implies
 a linear motion as it must be the case. We have  
$$
\lim_{t \to \infty} y(t) = \infty
$$
if and only if $n \le 1/3$,  i.e., for $n > 1/3$  the string comes to rest
with 
respect 
to the cosmic background after a finite time. A more stringent condition 
to be 
discussed is that the string comes to rest in a natural frame of a suitably 
chosen reference galaxy. This means  
$$
\lim_{t \to \infty} s(t) < \infty
$$
 and is fulfilled 
for $n \ge 1/2$. Therefore, the most interesting cases $n = 1/2$  (radiation
 model) and $n = 2/3$  (Einstein-de Sitter dust model) have the property that 
straight open strings come to rest after a sufficiently long time
independently 
of the initial conditions.

\bigskip

\noindent
{\it 2.   Example.}  Let $a(t) = e^{Ht}$, the inflationary  model, $H> 0$. 
Then $y(t) =e^{-3Ht}$, $
 s(t) =e^{-2Ht}$ . We have the same result as in the first example 
with $n \gg 1$.

\bigskip

\noindent
{\it 3.   Example.}  Let
$$
a(t) =  - t^{2/3}(1 + t^{-2} \cos mt) \, . 
$$
This background metric is from damped oscillations 
of a massive scalar field or, equivalently, from fourth order 
gravity   $L = R + m^{-2} \, R^2$. We get with eq. (10)
$$
      y(t) \approx   -1/t - ct^{-4} \sin mt \, , \quad 
    c = {\rm  const. \quad  for} \quad   t \to \infty \, , 
$$
which is only a minor modification of the result of the first example 
with $n = 2/3$, therefore, one should not expect a kind of resonance 
effect between 
the open string and a massive scalar field.

\subsection{The closed string}%3.2.

We insert $dy^2 + dz^2 = dr^2 + r^2 d\Phi^2$  into eq. (9) and use the 
space-like Killing vector $\pa/\pa\Phi$: its trajectories are circles, 
so we have in mind 
a closed string with radius $r$  moving (and eventually oscillating) in the 
expanding universe. The corresponding geodesic equation leads to
\bea
\frac{d}{d\lambda}
 (-a^4r^2\dot r) - a^2 r \dot t^2 + a^4 r (\dot x^2 + \dot r^2)=0 \, ,
\\
\frac{d}{d\lambda} (a^4 r^2 \dot x) =0 \,   ,
\\
\dot t^2 - a^2 (\dot x^2 + \dot r^2)= \frac{1}{a^2r^2}\, , 
\eea
where the dot denotes $d/d\lambda$, $\lambda$
 is the natural parameter along the 
geodesic. We are mainly interested in solutions not moving into 
the $x$-direction, to understand the oscillating behaviour. 
Inserting $x = 0$ into 
eqs. (11 - 13) and using the fact that always $\dot  t \ne  0$
 holds, eqs. (11-13) can be brought  into the form
\be%14
r \cdot r''a^3- 2 r r'a^4a'-r'^2a^3+3rr'a^2a'+a=0\, , 
\ee
where the dash denotes $d/dt$.

\bigskip

\noindent
{\it
4. Example.} Let $a = 1$, i.e., we have the flat Minkowski spacetime. 
Eq. (14) then reduces to
\be%15
rr''=  r'^2 - 1\, .
\ee
  The solution reads       
\be%(16)
           r(t) = r_0 \cos ((t - t_0)/r_0) \, , \quad
r'=     -\sin((t- t_0)/r_0) \, .
\ee
This is the oscillating solution for closed strings. At points $ t$ where
 $r = 0$, we have $ \vert  r' \vert = 1$, 
see eq. (16). These points are the often discussed cusp points 
of the string, where the interior string metric becomes singular and the 
motion approximates the velocity of light (cf. THOMPSON 1988).

\bigskip

Let us compare eq. (15) with the analogous equation for a positive definite 
back-ground metric. It reads
$$
rr''=  r'^2 + 1\, 
$$
and has solutions with cosh instead of cos. This is then the usual minimal 
surface taken up e.g. by a soap-bubble spanned between two circles.

\bigskip

\noindent
{\it 5.   Example.}  Let $a(t) = t^n$, then eq. (14) specializes to
\be%17
rr''t^{2n}
 - 2nrr'^3t^{4n-1} - r'^2t^{2n} + 3nrr't^{2n-1} + 1 = 0 \, .
\ee
This equation governs the radial motion of 
a circle-like closed string in a Friedmann background. The solutions can be 
hardly obtained by analytic methods.

\section{The string in an anisotropic Kasner background}%4

Now we take as background metric
\be%18
ds^2 =    dt^2 - a^2(t) dx^2 - b^2(t) dy^2 - c^2(t) dz^2
\ee
  and $\pa/\pa x$ as Killing vector. Astonishingly, the anisotropy has only 
a minor influence on the motion of the string, so we get mainly the same 
formulae  as in sct. 3.1.: the geodesic equations are
$$
\frac{d}{d\lambda} (a^2 b^2 \dot y) =0 
 \quad {\rm hence} \quad a^2 b^2 \dot y = M_1     
$$
$$
\frac{d}{d\lambda} (a^2 c^2 \dot z)  
= 0 \quad {\rm hence} \quad  a^2 c^2 \dot z
 = M_2        
$$
$$
  a^2 \dot t^2 - a^2  c^2 \dot z^2   = 1
$$
and can be integrated to yield for $M_1 M_2 \ne  0$
\be%19
     y(r) = \int \frac{dt}{b(t)
\sqrt{
\frac{a^2b^2}{M_1^2} +
\frac{M_2^2}{M_1^2}\cdot \frac{b^2}{c^2}
+1
}
}\, . 
\ee
The equation for $z(t)$ can be obtained from eq. (19) by interchanging
 $b \leftrightarrow c$ and $M_1 \leftrightarrow M_2$. 

\bigskip

For the Kasner 
metric we have eq. (18) with 
$a = t^p$, $b=t^q$, $c=t^r$, 
$p + q + r = p^2 + q^2 + r^2 = 1$. 
We get
$$
     y(r) = \int \frac{dt}{ t^q 
\sqrt{t^{2(p+q)}
+ t^{2(q-r)}
+1
}
}\, . 
$$
The behaviour for $t \to \infty$  in dependence
 of the values $p$, $q$, $r$ can be seen 
from this equation.

\section{The string in a gravitational wave}%5

As background metric we use the plane-wave ansatz
\be%20
ds^2
=    2dudv + p^2(u)dy^2 + q^2(u)dz^2 \, .
\ee
Eq. (20) is a solution of Einstein's vacuum equation if
\be%21
qd^2p/du^2 +pd^2q/du^2 =   0 \, .
\ee
Let us take $\pa /\pa y$  as Killing vector. Then the geodesic equations for 
the auxiliary metric are 
$$
\frac{d}{d\lambda} (p^2 \dot u) =0 
 \quad {\rm hence} \quad p^2  \dot u = M_1 \ne 0 \, ,    
$$
$$
\frac{d}{d\lambda} (p^2 q^2 \dot z)  = 0 \quad
 {\rm hence} \quad p^2 q^2 \dot z
 = M_2    \, ,     
$$
$$
 -2 p^2 \dot u \dot v - p^2 q^2 \dot z^2   = 1 \, .
$$
If we take $u$ as new independent variable (which is possible because 
of $\dot u \ne  0$) we get the solutions
\bea%22
     z(u) = \frac{M_2}{M_1} \int \frac{du'}{q^2(u')} \ ,   
\nonumber \\
 v(u) = \frac{- M^2_2}{2M^2_1} \int
\left( \frac{1}{q^2(u')} - p^2(u') \right) du'  \ .   
\eea
As an example let us take
\be%23
p(u) = \sin u \,  , \qquad
q(u) = \sinh u
\ee
which turns out 
to be a solution of eq. (21). Inserting (23) into eq. (22) we get
$$
z(u) = M \coth u\,  ,
\qquad 
v(u) = -M^2(4 \coth u + \sin 2u - 2u)/8 \, .
$$
Another Killing vector of eq. (20) is $\pa/\pa v$,
 but it is a null Killing vector and 
so the reduction used above does not work. But there exists a further 
non-null Killing vector of eq. (20). It reads
$$
k_i   = (0, y, H, 0), \qquad {\rm where} \qquad
 H = - \int p^{-2} du\,  . 
$$
By a coordinate 
transformation
$$
     u=t \qquad  y=w \cdot e^{-G} \qquad    G(u)= \int \frac{du}{p^2 H}
$$
$$
     z = k \qquad     v =  x + \frac{1}{2H} e^{-2G} (w^2-1)
$$
we get the form (6)
$$
ds^2 = e^{ -2G}p^2 dw^2
+  2 dx dt + 
 \frac{e^{ -2G}}{p^2H^2}
    dt^2 + q^2dk^2 
$$
and have to solve the equations
$$
\frac{d}{d\lambda} \left( p^2 q^2 e^{-2G} \dot k 
\right)=0 \, , \qquad  \frac{d}{d\lambda} \left( p^2 e^{-2G} \dot t 
\right)=0 \, , 
$$
$$
-p^2 q^2 e^{-2G} \dot k^2 - \frac{1}{H^2} e^{-4G}
\dot t^2 - 2p^2 e^{-2G} \dot x \dot t  
= 1\, . 
$$
For the special case $\dot k = 0$ one has finally
$$
x(t) = - \frac{1}{2} \int e^{-2G} \left(
 \frac{p^2}{M_2^2} + \frac{1}{H^2 p^2}
\right) dt  \, .
$$

\section{Discussion}%6

Let us suppose a spatially flat Friedmann model with scale 
factor $ a(t) = t^n$, and $n = 2/3$  (Einstein-de Sitter dust model) or 
$n = 1/2$ 
(radiation model). There we consider an open string which is only a little 
bit curved such that  the approximation of a straight string is applicable. 
At time $t = t_0 > 0$ we can prescribe place and initial velocity 
$v_0$ of the 
string arbitrarily and 
get for $t \to \infty$  the behaviour $s(t) \approx 
{\rm  const}  + M(v_0) t^{1 -2n}$  where 
$s$ denotes the natural distance from the origin. That means, for 
the cases $n = 2/3$ and $= 1/2$ we are interested in, the string comes to
rest 
for large values $t$ at a finite distance from the origin.

\bigskip

We compare this result with the analogous motion of a point-particle in 
the same background: in the same approximation we get 
$s(t) \approx {\rm 
const}  + M(v_0) t^{1-n}$,  i.e., 
$\vert s \vert \to \infty$  as $t \to \infty$, a totally other type of
motion. 
It should be mentioned that in the absence of gravity, i.e., $n = 0$, 
both motions 
are of the same type, but otherwise not.

\bigskip

The remaining calculations above indicate that the interaction of 
the motion of the string with scalar field oscillations (3. example), 
with anisotropy (sct. 4), and with gravitational waves (sct. 5) is quite
weak, 
we did not find any type of resonance effects.

\noindent 
{\it Acknowledgement.} We thank Dr. sc. U. KASPER 
and Dipl. Phys. O. HEINRICH for valuable discussions.

\section*{References}

\noindent 
BRANDENBERGER, R.: 1987, Int. J. Mod. Phys. A {\bf 2}, 77.

\noindent 
FROLOV, V., SEREBRIANY, E.: 1987, Phys. Rev. D {\bf 35}, 3779.

\noindent 
FROLOV, V., SKARZHINSKY, V., ZELNIKOV, A., 
HEINRICH, O.: 1988, PRE-ZIAP 88-14.

\noindent 
GEROCH, R., TRASCHEN, J.: 1987, Phys. Rev. D {\bf  36}, 1017.

\noindent 
GREEN, M., SCHWARZ, J., WITTEN, E.: 
1987, Superstring theory, Cambridge 1987.

\noindent 
NIELSEN, H., OLESEN, P.: 1973, Nucl. Phys. B {\bf  61}, 45.

\noindent 
STEIN-SCHABES, J., BURD, A.: 1988, Phys. Rev. D {\bf  37}, 1401.

\noindent 
THOMPSON, J.: 1988, Phys. Rev. D {\bf  37}, 283.

\noindent 
VILENKIN, A.: 1985, Phys. Rep. {\bf 121}, 263.

\noindent 
ZELDOVICH, Ya. B.: 1980, Mon. Not. R. Astron. Soc. {\bf  192}, 663.

\bigskip

\medskip

\noindent
Received 1988 July 28

\medskip

\bigskip

\noindent 
{\small {In this reprint done with the kind permission of the 
copyright owner 
we removed only obvious misprints of the original, which
was published in Astronomische Nachrichten:   
 Astron. Nachr. {\bf 310} (1989) Nr. 2, pages 103 - 108.

\bigskip

\medskip
\noindent 
  Authors's addresses  that time: 

\noindent
H.-J. Schmidt,  
Zentralinstitut f\"ur  Astrophysik der AdW der DDR, 
1591 Potsdam, R.-Luxemburg-Str. 17a

\medskip
\noindent
U. Semmelmann, 
Sektion Mathematik d. Humboldt--Universit\"at
Unter den Linden 8,
DDR-1086 Berlin}}

\end{document}